\documentstyle[aps,preprint,eqsecnum,epsfig,rotate]{revtex}

\newcommand{\beq}{\begin{eqnarray}}
\newcommand{\eeq}{\end{eqnarray}}

\newcommand{\ie}{{\it i.e.\ }}

\newcommand{\aphi}{\langle \phi \rangle}
\newcommand{\pbp}{\langle \bar{\psi} \psi \rangle}

\tightenlines
\begin{document}

\title{Finite density QCD via imaginary chemical potential}

\author{Massimo D'Elia$^a$ \footnote{E-mail: delia@ge.infn.it} and 
Maria-Paola Lombardo$^b$ \footnote{E-mail: lombardo@chimera.roma1.infn.it}}

\address{$^a$ Dipartimento di Fisica dell'Universit{\`a} di
    Genova and INFN, I-16146, Genova, Italy \\ $^b$
INFN, Sezione di Padova, and Laboratori Nazionali di Frascati, I-00044 Frascati, Italy }  

\maketitle

\begin{abstract}
We study QCD at nonzero temperature and baryon density
in the framework of the analytic continuation from imaginary
chemical potential.
We carry out simulations of QCD with four flavor of staggered fermions,
and reconstruct the phase diagram in the temperature--imaginary $\mu$ plane.
We consider ans\"atze for the analytic continuation of
the critical line and other observables motivated
both by theoretical considerations and mean field calculations
in four fermion models and random matrix theory.
We determine the critical line, and the analytic continuation
of the chiral condensate, up to $\mu_B \simeq 500$ MeV. 
The results are in qualitative agreement with the predictions
of model field theories, and consistent with a first order
chiral transition. The correlation between 
the chiral transition and the deconfinement transition
observed at $\mu=0$ persists at nonzero density.
\end{abstract}

\section{Introduction}

QCD at finite temperature and density is of fundamental
importance, both on purely theoretical and phenomenological grounds.
At high temperature asymptotic freedom will produce deconfinement
and chiral symmetry restoration, at high density a richer phase structure
and new phenomena have been predicted~\cite{Alford:1997zt}. 
In principle, the lattice formulation
provides a rigorous framework for the study of such phenomena.
In practice, however, the lattice regularization 
is usually combined with importance
sampling, which cannot be naively applied at nonzero baryon density,
where the quark determinant becomes complex~\cite{Kogut:1983ia}. 

It has been recently realized~\cite{Kogut:2002kk} 
that this problem can be circumvented
in the high $T$, low $\mu$ part of the QCD phase diagram where 
one can take advantage of physical fluctuations~\cite{Wilczek:1998ur}
Interesting physical information can be obtained by computing the
derivatives with respect to $\mu$ at zero chemical potential
and high temperature~\cite{susce,deriv,deriv1}. 
Fodor and Katz proposed an improved reweighting and applied
it to the study of the four~\cite{Fodor:2001au} and two plus one flavor 
model~\cite{Fodor:2001pe} . 
In refs.~\cite{Dag,HasTou92,alford}  
the  imaginary chemical potential approach 
was advocated and exploited in connection with the 
canonical formalism.
In ref.~\cite{mpl} it was proposed that the analytic continuation 
from imaginary chemical potential could be practical at 
high temperature, and the idea was tested  in the infinite coupling limit.
In refs.~\cite{hart1} ~\cite{Philipsen:2001ws}
the method was applied successfully to the dimensionally reduced model.
In ref.~\cite{deForcrand:2002ci}
 it was proposed that the critical line itself can be analytically
continued and results for two flavor of staggered fermions were presented.

In this work we  study QCD with four flavors of staggered 
fermions within the imaginary
chemical potential approach. Some of the results presented here have been
preliminarly reported in~\cite{D'Elia:2002pj}

In the next Section
we review the formalism and the method. In Section III  we reconstruct
the phase diagram in the temperature--imaginary chemical potential
plane. This is an interesting physical question by itself, it
is a mandatory step toward the reconstruction of the phase diagram
for real chemical potential, and provides some  guidance for
the analytic continuation. In Section IV we discuss a few aspects
of the analytic continuation and offer two examples
from model field theories: we will note there
that by considering $\mu^2$ rather than $\mu$, an 
analogy can be made between QCD at finite baryon density and ordinary
statistical systems in external fields. The remaining part of the paper
is devoted to numerical results at nonzero baryon density.
The critical line is presented Section V, including a first assessment
of the dependence on the number of flavors obtained combining the results
by de Forcrand and Philipsen~\cite{deForcrand:2002ci} 
with ours, and a cross check with the 
four flavor results by Fodor and Katz \cite{Fodor:2001pe}. 
The results for the chiral condensate are presented in Section VI.
In Section VII we discuss the nature of the chiral transition. Finally in 
Section VIII we summarize our results and give our conclusions.

\section{Formalism and method}

In the following we will briefly review the formulation of lattice QCD with
a non-zero chemical potential $\mu$ and the possible uses of working
with a purely imaginary $\mu$.
The zero density QCD partition function,
$Z(V,T) = {\rm Tr} \left( e^{-\frac{H_{\rm QCD}}{T}} \right) $,
with $H_{\rm QCD}$ the QCD Hamiltonian,
can be discretized on an euclidean lattice with finite temporal extent
$\tau = 1/T$
\beq
Z = \int  \left( {\cal D}U {\cal D}\psi {\cal D}\bar{\psi} \right)
e^{-\beta S_G[U]}e^{-S_F[U,\psi,\bar{\psi}]} = 
\int  \left( {\cal D}U \right) e^{-\beta S_G[U]} \det M[U]
\label{lattpf}
\eeq
where $U$ are the gauge link variables, $\psi$ and $\bar{\psi}$  are the fermionic variables, 
$S_G$ is the pure gauge action and  $S_F$ is the fermionic action which can be expressed in terms of the fermionic matrix
$M[U]$, $S_F = \bar{\psi} M[U] \psi$.

To describe QCD at finite density the grand canonical partition function,
$Z(V,T,\mu) = {\rm Tr} \left( e^{-\frac{H_{\rm QCD} - \mu N}{T}} \right) $,
where $N = \int d^3 x \psi^\dagger \psi$ is the quark number operator, can be used. 
The correct way to introduce
a finite chemical potential $\mu$ on the lattice~\cite{Kogut:1983ia} 
is to modify 
the temporal links appearing in the integrand in Eq.~(\ref{lattpf}) as follows:
\beq
U_t &\to& e^{a \mu} U_t \;\;\;\;\;\;\;\; {\rm (forward\; temporal\; link)} \nonumber \\
U_t^\dagger &\to& e^{- a \mu} U_t^\dagger \;\;\;\; {\rm (backward\; temporal\; link)}\; ,
\label{linktr}
\eeq
where $a$ is the lattice spacing.
$S_G$ is left invariant by this transformation but $\det M[U]$  
gets  a complex phase which makes importance sampling, and therefore
standard lattice MonteCarlo simulations, unfeasible.

The situation is different when the chemical potential is purely imaginary:
$U_t \to e^{i a \mu_I} U_t$, 
$U_t^\dagger \to e^{- i a \mu_I} U_t^\dagger$.
This is like adding a constant $U(1)$ background field to the original theory; 
$\det M[U]$ is again real and positive and simulations are as easy
as at $\mu = 0$.
The question then arises how simulations at imaginary chemical potential may be
of any help to get physical insight in finite density QCD.

One possibility is analytic continuation which should be practical at
relatively high temperature~\cite{mpl}.
$Z(V,T,\mu)$ is expected to be an analytical even function of $\mu$
away from phase transitions. For small enough $\mu$ one can write:
\beq
\log Z(\mu) &=& a_0 + a_2 \mu^2 + a_4 \mu^4 + O(\mu^6) \; \\
\log Z(\mu_I) &=& a_0 - a_2 \mu_I^2 + a_4 \mu_I^4 + O(\mu_I^6) \; .
\label{taylor}
\eeq
Simulations at small $\mu_I$ will thus allow a determination of the expansion 
coefficients for the free energy and, analogously, 
for other physical quantities, which can be cross-checked with those obtained 
by reweighting techniques (see ~\cite{glasgow,reveig,Crompton:2001ws} 
for further material on the reweighting approach).
This method is expected to be useful in the  high temperature regime, where 
the first coefficients should be sensibly different from zero; moreover
the region of interest for present experiments (RHIC, LHC) is that of 
high temperatures and small chemical potential, with $\mu / T \sim 0.1$.
This method has been already investigated in the strong coupling regime~\cite{mpl}, in the dimensionally reduced $3$--$d$ QCD theory~\cite{hart1},
and in full QCD with two flavours~\cite{deForcrand:2002ci}.
The Taylor expansion coefficients can also be measured as
derivatives with respect to $\mu$ at $\mu = 0$ ~\cite{susce,deriv,deriv1}.

$Z(V,T,i \mu_I)$ can also be used to reconstruct
the canonical partition function $Z(V,T,n)$ at fixed quark number $n$~\cite{roberge},
\ie at fixed density:
\beq
Z(V,T,n) &=& {\rm Tr} \left( ( e^{-\frac{H_{\rm QCD}}{T}} \delta(N - n) \right) = 
\frac{1}{2\pi} {\rm Tr} \left( e^{-\frac{H_{\rm QCD}}{T}} \int_0^{2\pi} 
{\rm d} \theta e^{i \theta (N - n)} \right) \nonumber \\
&=& \frac{1}{2\pi} \int_0^{2\pi}
{\rm d} \theta e^{- i \theta n} Z(V,T,i \theta T) \; .
\label{canonical}
\eeq
As $n$ grows, the factor
$e^{- i \theta n}$ oscillates more and more rapidly and the error in the 
numerical integration grows exponentially with $n$: this makes the application
of the method difficult especially at low temperatures where $Z(V,T,i \mu_I)$ 
depends very weakly on $\mu_I$.
The method has been applied in QCD ~\cite{HasTou92} and in
the 2--d Hubbard model~\cite{Dag}~\cite{alford},
where $Z(V,T,n)$ has been reconstructed up to $n = 6$
~\cite{alford}. 

The study of the phase structure of QCD in the $T$ -- $i \mu_I$ plane is also interesting by its own, as we will discuss in the next Section, and to 
understand the ranges of applicability of analytic continuation.

Results reported in the present paper refer to QCD with four degenerate 
staggered flavours of bare mass $a \cdot m = 0.05$ on a $16^4 \times 4$ 
lattice, where the  phase transition is expected at a critical coupling 
$\beta_c \simeq 5.04$~\cite{brown}. 
The algorithm used is the standard HMC algorithm.

\section{The phase diagram in the imaginary $\mu$ -- temperature space}

Let us write $Z(\theta) \equiv Z(V,T,i \theta T) = 
{\rm Tr} \left( e^{ i \theta N} e^{-\frac{H_{\rm QCD}}{T}} \right) $. Since
$N$ is a number operator, $Z(\theta)$ 
is clearly periodic in $\theta$ with period $2 \pi$; moreover a period 
$2 \pi /3$ is expected in the confined phase, where 
only physical states with $N$ multiple of 3 are present.
However it has been shown by Roberge and Weiss~\cite{roberge} that $Z(\theta)$ is always periodic 
$2 \pi /3$, for any physical temperature,
and that the only difference between the low $T$ and the 
high $T$ phase should be a smooth, analytic periodic behaviour at low $T$, 
as predicted from a strong coupling calculation, and a non-analytic 
periodic behaviour at high T with discontinuities 
in the first derivatives of the free energy at $\theta = 2 \pi /3 (k + 1/2)$, 
as predicted from a weak coupling calculation.
This suggests a very interesting scenario for the phase diagram of QCD in 
the $T$ -- $i \mu_I$ plane which needs confirmation by lattice calculations.

In order to get more insight into the phase structure of the theory, 
it is very useful
to consider the phase of the trace of the Polyakov loop, $P(\vec{x})$. 
Let us parametrize $P(\vec{x}) \equiv |P(\vec{x})| e^{i \phi}$,
and let $\aphi$ be the average value of the phase.
In the pure gauge theory the average Polyakov loop is non zero only in the
deconfined phase, where the center symmetry is spontaneously broken
and $\aphi = 2 k \pi / 3$, $k = -1,0,1$, \ie the Polyakov
loop effective potential is flat in the confined phase and develops 
three degenerate minima above the critical temperature.
In presence of dynamical fermions $P(\vec{x})$ enters explicitely
the fermionic determinant and $Z_3$ is broken: the effect of the determinant
is therefore like that of an external magnetic field which 
aligns the Polyakov loop along $\aphi = 0$. In the high temperature
phase the $Z_3$ degeneracy is lifted and $\aphi = 0$ is the true vacuum.

When $\mu_I \neq 0$, what enters the fermionic
determinant is $P(\vec{x}) e^{i \theta}$, $\theta \equiv \mu_I/T$, 
instead of $P(\vec{x})$.
Therefore the determinant now tends to
align $\aphi + \theta$ along zero, like 
a magnetic field pointing in the 
$- \theta$ direction. Hence one expects $\aphi = - \theta$ at
low temperatures.  At high temperatures the fermionic determinant
still lifts the $Z_3$ degeneracy, but it is $\theta$ that fixes which
is the vacuum. In particular for 
$ (k - 1/2) < \frac{3}{2 \pi} \theta < (k + 1/2)$ one expects
$\aphi \sim 2 k \pi / 3$ and 
$\theta =  2 (k + 1/2) \pi / 3$ should correspond to  
phase transitions from one $Z_3$ sector to the other. 

\begin{figure}
\vspace{2cm}
\epsfxsize=35pc 
\epsfbox{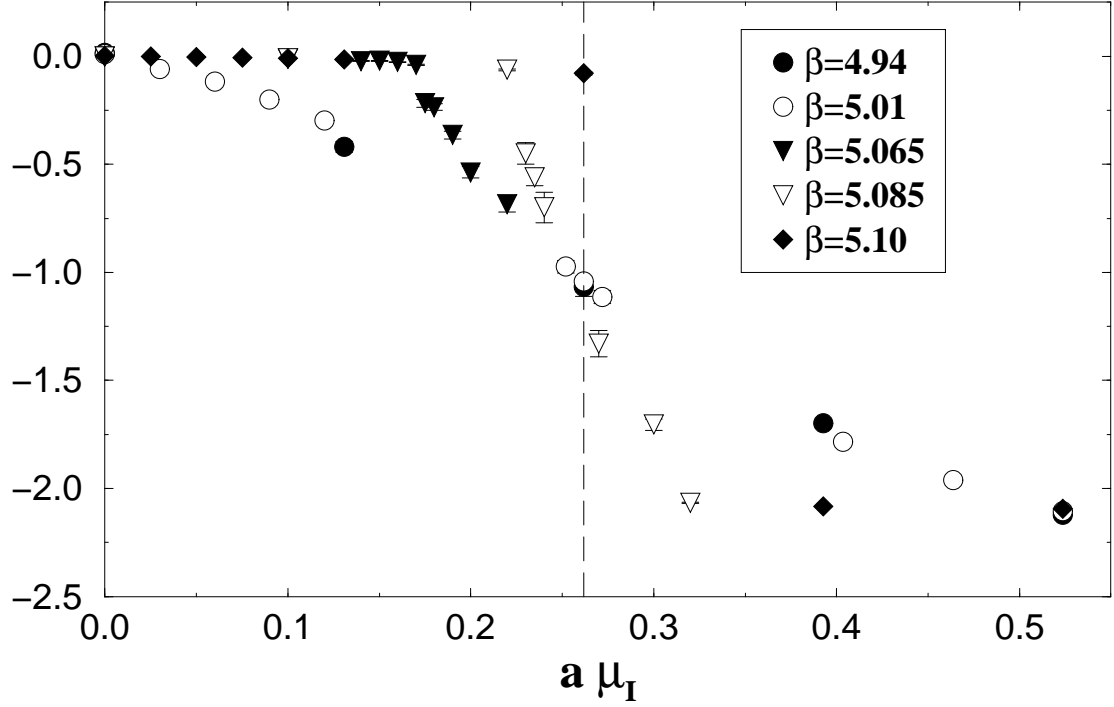}
\vspace{0.60cm}
\caption{Average value of the Polyakov loop phase as a function
of the imaginary chemical potential for different values of $\beta$.
The vertical dashed line corresponds to $\theta = \mu_I/T = \pi/3$.}
\label{phasefig}
\end{figure}

In Fig.~\ref{phasefig} we report our results for $\aphi$ versus the
imaginary chemical potential for different values of $\beta$. Since
$T = 1/(N_t a)$ and $N_t = 4$ in our case, we have $\theta = 4 a \mu_I$.
For $\beta = 4.94$ and $5.01$, which are below the critical $\beta$
at $\mu_I = 0$, $\beta_c(\mu_I=0) \equiv \beta_c \simeq 5.04$, one has 
$\aphi \simeq - \theta = - 4 a \mu_I$, \ie $\aphi$ is driven continously 
by the fermionic determinant. For $\beta = 5.10$, which is well
above $\beta_c$, we see that $\aphi \simeq 0$, almost independently of
$\mu_I$, as long as $\theta < \pi/3$, while for $\theta > \pi /3$
there is a sudden change to $\aphi \simeq - \pi/3$: we are clearly
crossing the Roberge-Weiss (RW) phase transition from one $Z_3$ sector
to the other. 
At intermediate values, $\beta = 5.065$ and $5.085$, $\aphi \simeq 0$
until a critical value of $a \mu_I$, where it starts moving almost 
linearly with $\mu_I$ crossing continuously the $Z_3$ boundary: in this case
there is no RW phase transition, but there is anyway a critical value
of $\mu_I$ after which $\aphi$ is no more constrained to be $\simeq 0$
and changes again lienarly with  $\theta$: as we will soon clarify,
this critical value of $\mu_I$ corresponds to the crossing of the 
chiral critical line, 
\ie the continuation in the $T$--$\mu_I$ plane of the chiral phase transition.

\begin{figure}
\vspace{2cm}
\epsfxsize=35pc 
\epsfbox{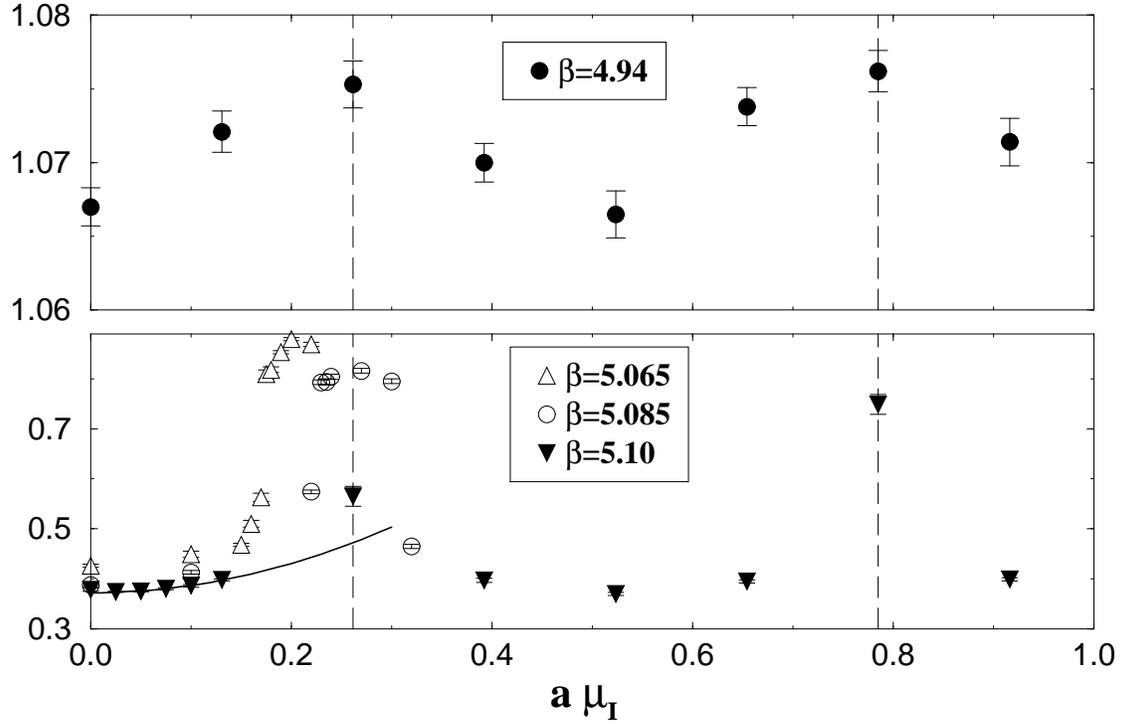}
\vspace{0.6cm}
\caption{Average value of the chiral condensate as a function
of the imaginary chemical potential for different values of $\beta$.
The vertical dashed lines correspond to $\theta = \mu_I/T = (2 k + 1) \pi/3$.
The continuous line in the lower picture is the result of a 
quadratic fit at small values of $a \mu_I$ obtained at $\beta = 5.10$.}
\label{psifig}
\end{figure}

We display our results for the chiral condensate in Fig.~\ref{psifig}. 
We expect a periodicity
with period $2\pi /3$ in terms of $\theta$. Moreover $\pbp$, like
the partition function, is an even function of $\mu_I$: 
this, combined with the periodicity, leads to
symmetry around all points $\theta = n \pi/3$,
with $n$ an integer number, for $\pbp$ as well as for the partition
function itself.
For $\beta < \beta_c$, $\pbp$ has a continuous dependence on 
$a \mu_I$ with the expected periodicity and symmetries.
For $\beta > \beta_c$ the correct periodicity and symmetries are
still observed but the dependence is less trivial. 
At $\beta = 5.065$ there is a critical value $a \mu_I \simeq 0.17$ for which 
the theory has a transition to a spontaneously broken chiral
symmetry phase: we are clearly going through the chiral critical line. 
The same happens for $\beta = 5.085$ at $a \mu_I \simeq 0.22$: in this 
case we have proceeded further, observing also the transition back
to a chirally restored phase at $a \mu_I \simeq 0.30$, which is, correctly,  
the symmetric point with respect to $\theta = \pi / 3$.
At $\beta = 5.10$ we never cross, when moving in $\mu_I$, 
the chiral critical line, but only  RW critical lines~\footnote{
Error bars for the determinations at $\beta = 5.10$ and 
on the critical lines ($\theta = \pi/3$ and $\theta = \pi$) 
are probably underestimated.}.

Besides the sets of runs at fixed $\beta$ and variable $\mu_I$, 
we have also performed runs at fixed $\mu_I$ and variable $\beta$ to 
look for other locations of the chiral line in the $T$--$\mu_I$ plane.
In every case the location of the phase transition has not been determined
by looking at susceptibilities, since our statistics for each single
run were rather
poor to this aim (of the order of 1000 molecular dynamics time units), 
but rather by looking at sharp changes of various physical quantities, 
among which the chiral condensate or the Polyakov loop. In each case 
sharp drops or jumps have been observed, allowing quite precise determinations
of the transition point and suggesting the first order nature
of the phase transition also at $\mu_I \neq 0$.
The drop of the condensate is always
coincident with the sharp jump of the Polyakov loop, also at 
$\mu_I \neq 0$, suggesting that the coincidence of chiral symmetry
restoration and deconfinement holds true also at $\mu_I \neq 0$.
A summary of all our determinations of the chiral critical line is reported
in Table~\ref{crititable}.

\begin{figure}
\vspace{2cm}
\epsfxsize=35pc 
\epsfbox{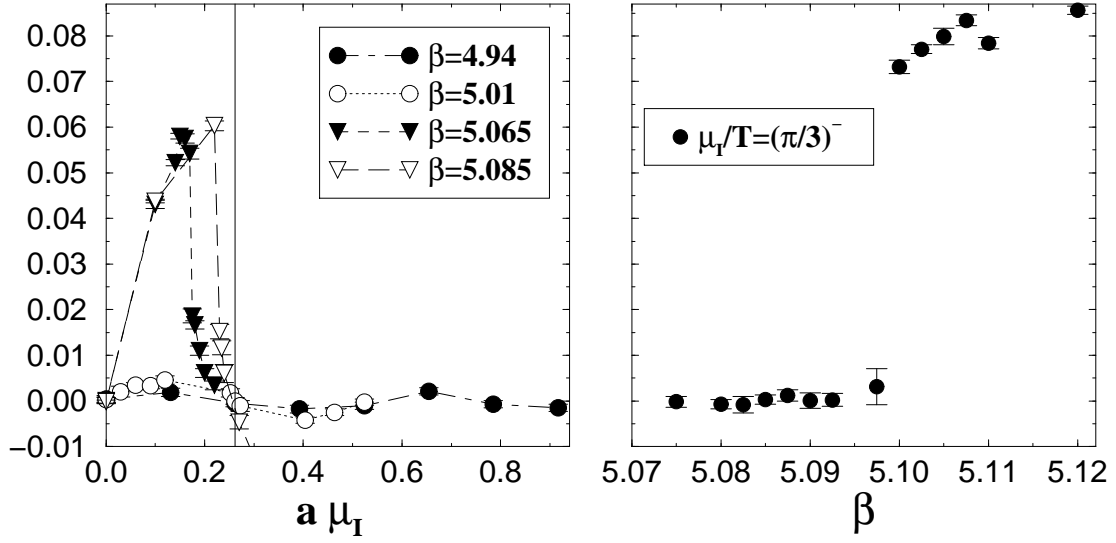}
\vspace{1.00cm}
\caption{Imaginary part of the barion density as a function of
$\mu_I$ for different values of $\beta$ (left--hand side), and 
as a function of $\beta$ at $\theta = \mu_I/T =  \frac{\pi}{3}^-$
(right-hand side).}
\label{barfig}
\end{figure}

\begin{figure}
\vspace{0cm}
\epsfxsize=35pc 
\epsfbox{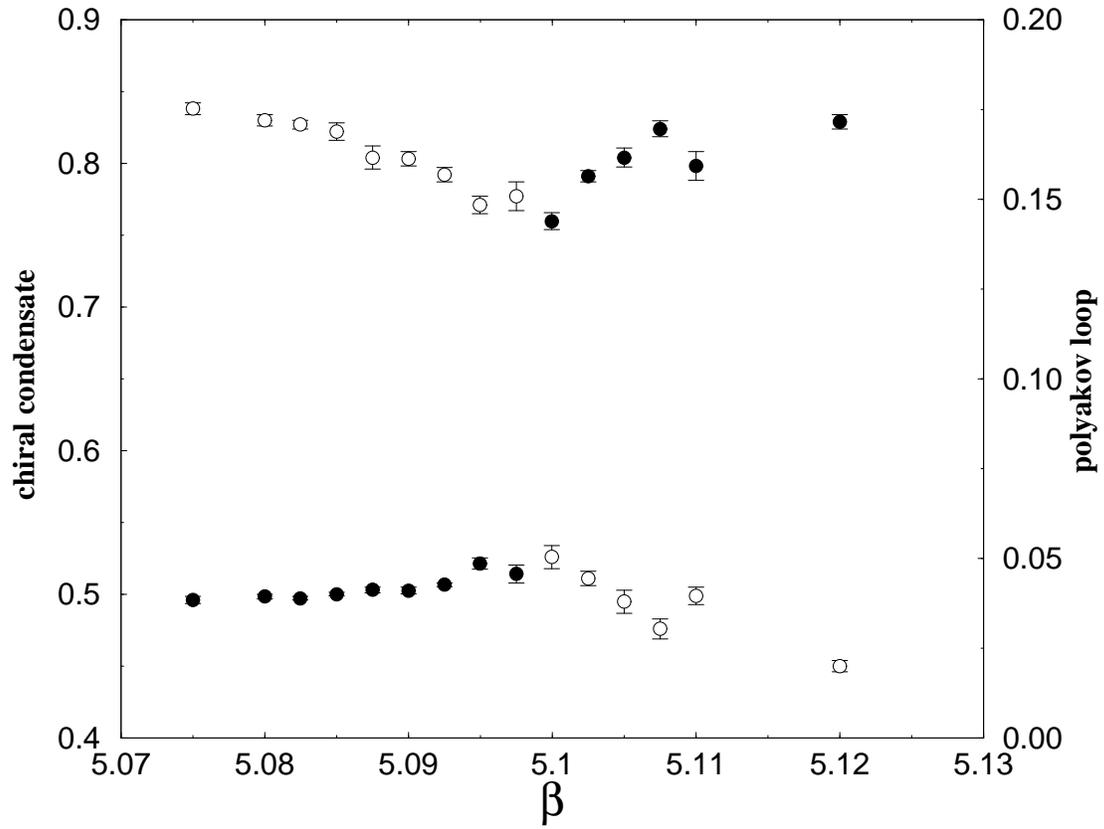}
\vspace{0.60cm}
\caption{Chiral condensate (white circles) and absolute value of the 
Polyakov loop (black circles) as
a function of $\beta$ for $\theta = \pi/3$. The sharp changes of the 
two quantities coincide with the location of endpoint of the RW critical
line.}  
\label{polpsi}
\end{figure}

It is interesting to illustrate in more details the determination of the 
endpoint of the RW critical line, $\beta_E = 5.097(2)$. We have performed
a simulation at exactly $\theta = \pi/3$, starting thermalization 
from a zero field configuration: we thus drive the system 
to one side of the RW critical line (assuming that the line is there), 
\ie on  the border of one $Z_3$ sector, 
since on the $16^3 \times 4$ lattice it is already practically unfeasible 
to flip through the RW line in reasonable simulation time. 
Let us now consider the baryon density, 
$\langle b \rangle = \frac{T}{V} \frac{\partial}{\partial \mu} \ln Z$:
it is an odd function of $\mu$, since $Z$ is an even function. Therefore, 
for an imaginary chemical potential, $\langle b \rangle$
is also purely imaginary and an odd function of $\mu_I$. This, combined
with the periodicity in $\mu_I$, leads to the expectation that
$\langle b \rangle (\theta = \frac{\pi}{3}^-) = - \langle b \rangle 
(\theta = \frac{\pi}{3}^+)$.
The last relation clearly implies that $\langle b \rangle = 0$ at 
$\theta = \pi/3$, unless $\langle b \rangle$ is not continuous on that
point. Thus a non-zero value of $\langle b \rangle$ at 
$\theta = \frac{\pi}{3}^-$ implies the presence of the Roberge-Weiss
critical line. 
On the right hand side of Fig.~\ref{barfig} the imaginary part of
$\langle b \rangle$ at $\theta = \frac{\pi}{3}^-$ is plotted as a function
of $\beta$: one can clearly see a transition from a zero to a non-zero
expectation value, which permits the determination of $\beta_E$.
We have verified that at $\beta_E$ also the chiral condensate and the
polyakov loop have a sharp change, as we show in Fig.~\ref{polpsi}, 
and this implies that at this point, for $\theta = \pi/3$, we 
also meet the chiral critical line, so that the RW
critical line ends on the chiral critical line. 
On the left hand side of Fig.~\ref{barfig} we present instead 
the imaginary part of $\langle b \rangle$ as a function of
$\mu_I$ for different values of $\beta < \beta_E$: in this case
$\langle b \rangle$ is always zero and continuous 
at $\theta = \frac{\pi}{3}$, but
it is interesting to note how it starts developing the discontinuity
as $\beta \to \beta_E$.

\begin{figure}
\vspace{2cm}
\epsfxsize=35pc 
\epsfbox{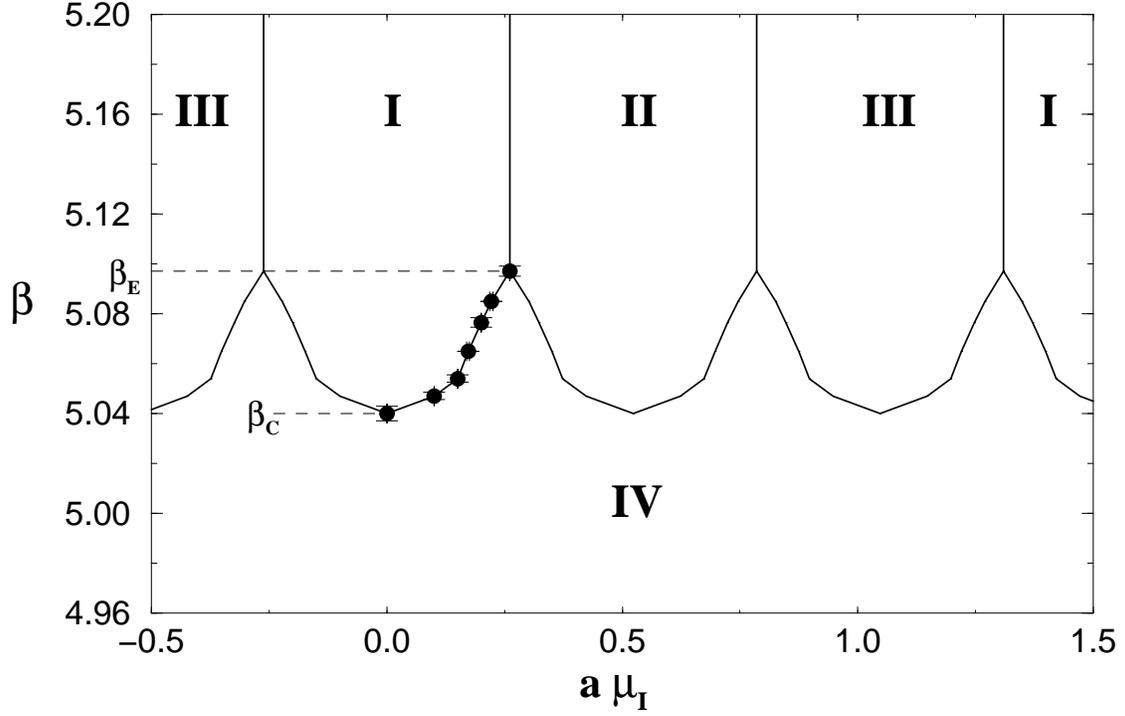}
\vspace{0.60cm}
\caption{Sketch of the phase diagram in the $\mu_I$--$\beta$ plane.
The filled cirles represents direct determinations of the chiral critical 
line location from our simulations. The rest of the chiral line 
has been obtained by interpolation and by exploiting
the symmetries of the partition function.}
\label{diagfig}
\end{figure}

We present, in Fig.~\ref{diagfig}, a sketch of the phase diagram
in the $\beta$--$\mu_I$ plane, as emerges from our data
and by exploiting the above mentioned symmetries.
We can distinguish a region where chiral symmetry is spontaneously 
broken (indicated as IV in the figure) and three regions (I, II and III),
which correspond to different $Z_3$ sectors and repeat periodically, 
where chiral symmetry is restored. The chiral critical line
separates region IV from other regions, while the RW critical lines
separate regions I,II and III among themselves.
As we have noticed above, the sharp drops fo the chiral condensate
at the transition points suggest that the chiral critical line is first
order also at $\mu_I \neq 0$, so that we expect all regions to be 
separeted by first order critical lines.

\section{From imaginary to real $\mu$  and the critical line in the 
$T,\mu^2$ plane}

We will concern ourselves with the properties of the critical line,
as well with the $\mu$ dependence of a number of physical 
observables. 

To this end one needs to analytically continue the results
from purely imaginary to real chemical potential:
generically speaking,  one deals with a function of a purely
imaginary variable, continues it to the entire complex plane, and finally
takes the limit $\epsilon \to 0$ of $f(x + i \epsilon)$, $x$ and $\epsilon$
being real variables. General arguments guarantees that the analytical
continuation is unique in the analyticity domain of the function.
In practice,  $f(x)$ is unknown and has to be approximated by some series
expansion or suitable ansatz. In either case,
one has to pay attention to the fact that,
by modifying the original expression by a non--leading term,
the difference in physical quantities is still non leading.
This can be complicated, and often mathematical arguments need to be
supplemented by some physical insight ~\cite{baym}, 
even in relatively simple cases ~\cite{evans}. 

The critical line itself can be analytically continued from
imaginary to real chemical potential: a nice argument together
with an application to the two flavor model has been given
in \cite{deForcrand:2002ci}. 
The analytical continuation of the critical line is
easily discussed by considering that 
$\cal Z(\mu) $, an even function of $\mu$,  is \underline
{real valued} for either real and purely imaginary $\mu$ : for real
$\mu$,  the imaginary part of the determinant cancels out in the statistical
ensemble, and it is even possible to cancel it exactly on
a finite number of configurations by considering the appropriate symmetry
transformation \cite{Alles:2002wh,Ambjorn:2002pz}; 
for imaginary chemical potential,
the determinant itself is real, yielding authomatically
a real partition function. For a complex $\mu$, $\cal Z(\mu) $ is,
in general,  complex. 

We can  map the complex $\mu$ plane onto the complex
$\mu^2$ plane, and consider ${\cal Z}(\mu^2)$ (on a finite lattice this is
an exact polynomial).  
Then, $\cal {Z}$ is real valued on the real $\mu^2$ axis, complex
elsewhere : the situation is  analogous to  e.g. the
partition function as a function of a magnetic field, which becomes
complex as soon as the external field becomes complex, and the physical
domain (real partition function) is associated with real values
of the couplings.
The critical behavior of the system is then dictated by the zeroes
of the partition function (Lee-Yang zeros) in the complex $\mu^2$ plane.
The locus of the Lee Yang zeros is thought to be associated
to a general surface of phase separation \cite{janke}, and
phase transition points, for each value of the temperature, 
are associated with the Lee-Yang edge building up in the
infinite volume limit, thus defining  a curve in the $T, \mu^2$ plane.

This simple reasoning shows that it is very natural to think of the 
critical line as a \underline{smooth} function
$T(\mu^2)$, making it natural the analytic
continuation from positive to negative
$\mu^2$ values. Indeed, 
experience with statistical models shows that not only
the critical line,  but also the critical exponents 
are smooth functions of the couplings \cite{potts} (aside of course
from endpoints, bifurcation points, etc.). Hence,  they can
be safely expanded, either via Taylor expansion or
a suitable ansatz. In particular,
$\mu^2_c=0$ has no special character: it
is just the point where the Lee Yang edge hits the real axis where
$T=T_c$. 

Clearly the analytic continuation from imaginary chemical potential 
is practical when the critical line is smooth, and a few coefficients
suffice to describe it.
It is then of some interest to discuss  examples whose critical
lines are exactly computable:
we will indeed find that a second order expression in $T$ and $\mu$
 well approximates the critical
line \underline{over a large $\mu$ interval}.

\subsection{The Gross-Neveu model}

The Gross-Neveu model in three dimensions  is
interacting, renormalizable and can be chosen with
the same global symmetries as those of QCD
which, when spontaneously broken at strong coupling, produce
Goldstone particles and dynamical mass generation. 
As such, the Gross-Neveu model (as well as other four fermion
models) can provide some  guidance to the understanding
of the QCD critical behavior 
(see e.g.refs. \cite{Hands:2001jn,Hands:1998jg,kli}).

The critical line for the three dimensional GN model
was calculated in ref.~\cite{Hands:1992ck} and reads
\begin{equation}
1 - \mu /\Sigma_0 = 2 T /\Sigma_0 \ln (1 + e^{-\mu/T})
\label{eq:eosgn}
\end{equation}
where $\Sigma_0$ is the order parameter in the normal
phase. Setting $\mu=0$ in the above equation gives
the critical temperature at zero chemical potential, 
$T_c(\mu=0) = \Sigma_0/2 \ln 2 \simeq .72 \Sigma_0$
 
Expanding now $\ln (1+e^{(-x)}) \simeq \ln 2 - 1/2 x + 1/ 8 x^2$,
and eliminating $\Sigma_0$ in favour of $T_c$, we get
\begin{equation}
(T - 1/2 T_c)^2 + \mu^2/(8 \ln 2) = T_c^2 / 4
\end{equation}
It is easy to check that this expression approximates very well the exact 
result ~\ref{eq:eosgn}, and then that a second order expression in
$\mu$ is a good approximation to the critical line in this model.

\subsection{Random Matrix Theories}

As it is well known (see e.g. \cite{RMT1}), 
there is a remarkable relation
between the symmetry breaking classes of QCD and the
classification of chiral Random matrix Ensembles.

For QCD with fermions in the complex representation (i.e. $N_c > 2$,
fundamental fermions) with pattern of SSB 
$SU(N_f)_R \times SU(N_f)_L \rightarrow SU(N_f)$, the corresponding
RMT is chiral unitary, $\beta =2$ in the Dyson representation.
On the lattice,  staggered fermions have unusual patterns of 
$\chi SB$: all real and pseudoreal representations
are swapped. However, for complex representations, the corresponding
RMT ensemble remains chiral unitary \cite{RMT1}.
The critical line in the $T, \mu$ plane for this ensemble
derived in ref. \cite{RMT}:
\begin{equation}
(\mu^2 + T^2)^2 + \mu^2 - T^2 = 0
\label{eq:eosrmt}
\end{equation}
is thus valid both on the lattice and in the continuum. 
Expanding it to $O(\mu^2$) we obtain
\begin{equation}
T^2 = T_c^2 - 3 \mu^2  
\end{equation}
and by comparison with the exact result~\ref{eq:eosrmt}, 
we note that this expression
describes well the critical line basically till its
endpoint.

\section{The critical line for the four flavor model}

In Section III we have presented our measurements of the
critical points in the temperature imaginary chemical potential
plane, see again Table~\ref{crititable}. 
Here we use those data to reconstruct the
critical line for real chemical potential.
We used both least
$\chi^2$ and least squares fits to a second order
polynomial,  and we studied the effect of a fourth
order term.

A significant sample of 
the results of the least $\chi^2$   and least squares fits to
a second order polynomials
\begin{equation}
\beta_c(\mu) = a + b \mu^2 
\end{equation}
are collected in  Table~\ref{crititable2} and Table~\ref{tablefitcrit}.

The weight of each data point for the least square fits is zero or
one, i.e. we either include or discard it, and the quality of our fits
-for different $\mu$ ranges indicated in the Table- 
is measured by the squared sum of residuals, to be compared with
$ \sum (\Delta \beta)^2 $. Some of the fits (whose results we omit) 
included a linear term, and we confirm that it is compatible with zero,
as it should on symmetry grounds.  The 
least $\chi^2$ fits have been performed by discarding one point
at the time.

From the least square polynomial fits we obtain
\begin{equation}
\beta_c = 5.038(2) + 0.94(7)\mu^2_c
\end{equation}
and from the minimum $\chi^2$  polynomial fits:
\begin{equation}
\beta_c = 5.036(2)(2) + 0.98(3)(6)\mu^2_c 
\end{equation}
The central values are the average of the results, the
first error is the mean (absolute) deviation, the
second one is the maximum error measured in individual fits.
We combine all of these estimates to quote as our final result for
the critical line in lattice units:
\begin{equation}
\beta = 5.037(3) + 0.96(10)\mu^2 + O(\mu^4)
\label{fig:latres}
\end{equation}

The  continuation to real chemical potential  
is shown in Fig. \ref{fig:reale_uno}. The dotted lines
are drawn in correspondence of the
central values of the fit parameters 
\ref{fig:latres} plus and minus the quoted errors.
The dispersion remains reasonable in a fair interval of chemical
potentials (  $\mu = .2$ corresponds, roughly, to a baryochemical potential of
about 370 Mev and $\mu = .3$ to about 516 Mev.) Moreover,
in the same interval, we note a nice agreement between our results
and those obtained by Fodor and Katz via their improved 
reweighting~\cite{Fodor:2001au}
in the same four flavor model and with the same quark mass $m_q = 0.05$. 
Obviously our results, being  uncorrelated, look less regular.

The line obtained via improved reweighting  changes concavity
around $\mu \simeq .4$. This is not easy
to understand from a physical point of view, 
but, from a purely numerical perspective,
this behavior could be reproduced by a negative fourth order
coefficient in the expansion of the critical line.
It would then be desirable to place  at least  bounds on the coefficient of 
the fourth order term, but, on the other hand,
as the quadratic fits are satisfactory, this is not an easy  task.
To get a feeling
of the effect of a fourth order term we constrained the constant
and the second order term to leave the fourth order coefficient
as a variable.  The results are given in Fig.~\ref{fig:reale_due}. 
The solid line corresponds to
the result of a second order fit up to $\mu = .2$.
The dotted line is the correction to that result induced by
a fourth order correction, which is very small.
More sizable corrections are induced  by
constraining the second order coefficient to its extreme
value (as inferred from the errorbars) and then fitting the
fourth order term as a free parameter.

From Fig.~\ref{fig:reale_uno} and Fig.~\ref{fig:reale_due}
we conclude that
the imaginary chemical potential approach provides  a
safe estimate for the critical line up
to $\mu \simeq .3$ (corresponding to a baryochemical
potential of $\simeq$ 500 MeV) before loosing accuracy, and that
within this interval imaginary chemical potential and
reweighting give consistent results for the critical line.
The loss of accuracy for the
imaginary chemical potential results 
is mostly due to the influence  of the fourth order
term: we emphasize anyway that the results obtained with 
a fourth order term are much less accurate, however consistent
with the ones coming from a second order polynomial.
To reduce the error band we would need better data for the critical line
at imaginary chemical potential. 

To convert to physical units we need the lattice spacing as
a function of the coupling. We used as an input the lattice spacing
measured at $\beta = 5.10$ ~\cite{Alles:2000cg} 
to fix the scale in the two loop
$\beta$ function. We have verified that the ratio $r_{meas}$ of the
scales taken at the two extrema of the interval of
interest [5.04:5.10]  measured in ~\cite{Alles:2000cg} 
agrees within a few percent
with the ratio $r_\beta$ coming from the two loop $\beta$ function.
Clearly the uncertainty induced by the
interpolation and/or by the choice of the 
$\beta$ used as an input is less than the numerical errors on the
lattice results.

In Fig. \ref{fig:reale_tre} the external band  show the results of the
conversion to physical units by use of the two loop
$\beta$ function, the errors being those induced
by the analytic continuation. In the same plot we draw the
ellipse arc
\begin{equation}
T = \sqrt{( T_c^2 - k \mu^2)}
\end{equation}
which turns out to be a nice approximation to the data in physical units,
with $k = 0.021$ (this just  the result of the
fit in the interval [0:400 MeV] to the central values of our critical
line). Other functional forms would work as well, and,
in particular, the  parabola 
$T  = T_c (1 - 0.5 k \mu^2 /T_c^2)$
obviously provides  a good approximation to the data, given the smallness
of $(k/T_c)^2$.

The results by De Forcrand and Philipsen \cite{deForcrand:2002ci}
 combined with ours
allow an  assessment of the flavor dependence of the
critical line in QCD: this is done in Fig.~\ref{fig:reale_quattro} in the
form of a scale invariant plot. We omit the errorbars for the sake of clarity,
and just remind the reader that the errors -- both for 
the two and the four flavor model -- are reasonably small up to
$\mu_B \simeq 500$. We see  that
the transition in  four flavor QCD lies consistently below 
that of the two flavor theory, and that this effect increases with
increasing density. It
looks as if the production of real fermions further favors the
phase transition, which is indeed the expected behaviour
(see e.g. ~\cite{Karsch:2001cy}).

\section{Chiral condensate}

Taylor expansion and Fourier decomposition are natural parametrisation
for our observables.
In particular, the  analysis of the phase diagram in the temperature-imaginary 
chemical potential plane suggests
to use Fourier analysis for $T \le T_c$. Moreover, the fugacity expansion
\begin{equation}
{\cal Z} =  \sum c_n e^{3 \mu n N_t} 
\end{equation}
hints that this would be easier in the cold phase, 
at stronger coupling, where a few coefficients might suffice.
At high temperature, in the weak coupling regime, on the other hand,
perturbation theory might serve as a guidance, 
suggesting that the first few terms of the
Taylor expansion might be adequate in
a wider range of chemical potentials.  

As the chiral condensate is an even function
of the chemical potential, its Fourier decomposition reads:
\begin{equation}
\langle \bar \psi \psi \rangle = \sum_n  a_F^{n} \cos (n N_t N_c \mu_I)
\end{equation}
which is easily continued to real chemical potential
\begin{equation}
\langle \bar \psi \psi \rangle = \sum_n  a_F^{n} \cosh (n N_t N_c \mu_R)
\label{eq:pbpcosh}
\end{equation}

In our Fourier analysis of the chiral condensate results 
we limit ourselves to $n=0,1,2$ and we assess the validity
of the fits via both the value of the $\chi^2$ and the stability
of  $a_F^{0}$ and $a_F^{1}$ given by one and two cosine fits.
We summarize the results of the Fourier analysis of
the chiral condensate results in Table~\ref{cosfitpbp}.

The Fourier analysis
turns out to be satisfactory at $\beta = 5.01$ and $\beta = 5.03$
where -as discussed in Section III and shown in Fig. 2 above- 
the chiral condensate is a continuous function of $\beta$. 
One cosine fit is actually enough to describe
our data at $\beta = 5.01$ and $\beta = 5.03$, 
with the current statistical accuracy: adding a
term  $\cos (2 N_t N_c \mu) = \cos (24 \mu)$ in the expansion 
does not modify the value of
the coefficients $a_F^{0}$ and $a_F^{1}$,
and does not particularly  improve the $\chi^2$. 
The term $\cos (24 \mu)$ is anyway needed 
in order  to asses the errors on the analytic continuation to
real $\mu_B$, as we will discuss below.

At $\beta = 5.10$ (in the fits we discarded of course the points corresponding
to the RW discontinuity) we know (see again Section III and Fig. 2) 
that the periodicity is no longer smooth. Indeed, the
values of the first two Fourier coefficients depend on the type of the fit 
(two or three parameters).

Next, we have  considered polynomial fits.
In these fits  we exploited the symmetries and the periodicity 
of the model to improve the statistical accuracy: all
of the results were traslated, or symmetrised, to the first half period.
For a quick comparison with the Fourier analysis 
we use a polynomial fit of the form 
\begin{equation}
\langle \bar \psi \psi \rangle  = a^P_0 - 72 a^P_2 \mu^2 + 864 a^P_4 \mu^4
\end{equation}
When one cosine fit is adequate, the Taylor expansion would give
$a^P_0 = a_F^1 + b_F^1$, and $a^P_2 = a^P_4$. At $\beta = 5.01$ and
$\beta = 5.03$ this is indeed the case, within our largish errors.
For $\beta = 5.10$ a second order Taylor expansion is adequate 
at $\mu < 0.2$, while a fourth order term does not 
substantially improve the behavior. As discussed above, the quality of the 
polynomial fits should improve at higher temperature, closer to the 
perturbative regime, where a second order polynomial should become exact.
Results for the polynomial fits are collected in Table~ \ref{polfitpbp}.

We can now continue the results of the Fourier analysis 
to real chemical potential.
Fig. 10
  shows the behavior
of the chiral condensate as a function of real chemical potential
for $\beta = 5.01$ and $\beta = 5.03$, using \ref{eq:pbpcosh}
for both one and two cosine fits. We show the errorband induced by
the fits with a $\cos(24 \mu)$ term in a large chemical potential
range. The symbols (triangles and squares) are instead plotted
only in  the broken phase, i.e. for   $ \mu < \mu_c(\beta)$ 
(see next Section for more on this point).

We have checked that the analytic continuation to real chemical
potential of the results of the polynomial fits:
\begin{equation}
\langle \bar \psi \psi \rangle  = a^P + 72 a^P_2 \mu^2 + 864 a^P_4 \mu^4
\end{equation}
is affected by comparable, if not smaller, systematic errors
as the ones observed with the Fourier parametrization.

As a final comment on the polynomial 
fit results, we consider the temperature
dependence of $a^P_2$. The Maxwell relation
\begin{equation}
\frac {\partial J_0}{\partial m} = 
\frac {\partial \langle \bar \psi \psi \rangle}{\partial \mu}
\end{equation}
shows that $a^P_2$ is the derivative with respect to
the quark mass of the quark number susceptibility
(by taking the $\mu$ derivative of either sides of the
relation above). These results suggest that such derivatives
-in lattice units- do not change much with temperature, in contrast 
with the quark number susceptibility itself: the 
results for the number density (we postpone a discussion
of thermodynamics to another publication and we just quote the
central values) are:
\begin{eqnarray}
J_0 = 0.055 \mu  + 1.30 \mu^3 \\
J_0 = 0.073 \mu + 1.44 \mu ^3  \\
J_0 = 0.455 \mu  + 1.39 \mu ^3 
\end{eqnarray}
at $\beta = 5.01, 5.03, 5.10$ from top to bottom. The coefficient
of the first term is the quark number susceptibility in lattice
units, which increases rapidly in the plasma phase as it should.

\section{Nature  of the chiral phase transtition}

We have discussed in Section II 
the interrelation between chiral
condensate and Polyakov loop at least up to $\mu = .2$. 
Obviously, the observed correlation
holds true at  real chemical potential as well: 
we can consider the difference between the critical
coupling for the chiral condensate and the Polyakov loop: 
$\beta_c^\chi(\mu) - \beta_c^P(\mu)$. 
Our results  suggest that 
$\beta_c^\chi(\mu) - \beta_c^P(\mu) = 0$ 
at imaginary $\mu$ in a nonzero interval. It
should then remain zero also at real chemical potential. 
We can then conclude that, within the present accuracy, the 
nearly coincidence of chiral and deconfining transition
persists at nonzero baryon density in a four flavor model.

As for the order of the phase transition, 
let us consider again Fig. 10
where,  together with the errorband in a larger $\mu$ interval, we have 
plotted , as triangles and squares, the results from one and two cosine fits, 
for $\mu < \mu_c(\beta)$, inferred from the results
of Section III above. 
The behavior shown in Fig. 10
is then consistent  with a first order  transition.

\section{Summary}

We have studied the phase diagram of four flavor QCD in the
imaginary chemical potential -- temperature plane. 
We have measured the location of the endpoint of the
Roberge Weiss line and inferred that the RW line ends on the 
chiral critical line.

We have continued our results to real chemical potential.
The critical line in lattice units reads
$$\beta_c = 5.037(3) - 0.96(10)\mu^2 + O(\mu^4)$$
which, in physical units, is well described by 
$$T_c = \sqrt{(T_c^2 - 0.021 \mu^2)}$$ or, equivalently,
by
$$T_c = T_c \left( 1 - \frac{0.021}{T_c^2} \mu^2 \right)$$
over a large interval of chemical potentials. It is 
of some interest to notice that this feature is observed
also  in simple models, where we have found that 
the critical line is well approximated  by a
second  order polinomial  up to $\mu_{quark} \simeq T$.

We have found a good agreement with the results by Fodor
and Katz~\cite{Fodor:2001au} up to $\mu \simeq 400$ MeV. 
Our results are still
compatible with theirs, within largish errors, up to $\mu \simeq 500$.
Beyond that $\mu$ value our results do not have much statistical
significance and a comparison is no longer meaningful. 
We emphasize that the main source of uncertainty on our
results is statistical, and it is associated to a poor knowledge
of the fourth order coefficient, which appears to be very small.

We have studied the flavor dependence of the results
by comparing our findings for the critical line of four
flavor QCD with those obtained  by 
De Forcrand and Philipsen~\cite{deForcrand:2002ci}
in the two flavor model,
and found it to be that expected on physical grounds. 

We have studied the character of the chiral transition. We have
found indications that it remains correlated with the transition associated
with the Polyakov loop.  The dependence of the chiral condensate
on the chemical potential is consistent with a first order 
transition.

\section*{Acknowledgments}
We thank Ph. De Forcrand, Z. Fodor, F. Karsch, H. Neuberger, O.Philipsen, 
H. Satz,  M. Stephanov, B. Svetitsky, D. Toublan
 for useful discussions and comments.
This work has been partially supported by MIUR and ECT$^*$.
We thank the computer center of ENEA for providing us with time on
their QUADRICS machines.

\begin{table}
\setlength{\tabcolsep}{3.0pc}
\caption{Locations of the chiral critical line}
\vspace{0.3cm}
\label{crititable}
\begin{tabular}{llllll}
& & $\mu_I$ & $\beta_c$ & & \\
\hline
& & & & & \\
& & 0.00    &   5.0400(30) & & \\ 
& & 0.10    &   5.0470(15) & & \\ 
& & 0.15    &   5.0540(15) & & \\
& & 0.173(3)   &   5.0650 & & \\
& & 0.20    &   5.0765(20) & & \\
& & 0.222(3)   &   5.0850  & & \\
& & 0.2617994 & 5.0970(20)  & & \\
\end{tabular}
\end{table}

\begin{table}
\caption{Results of least $\chi^2$  quadratic fits to the critical line}
\setlength{\tabcolsep}{1.1pc}
\vspace{0.3cm}
\label{crititable2}
\begin{tabular}{crrrr}
 Discarded $\mu$  & a & b  & $\chi^2$/d.o.f. \\
\hline
  0.00&   5.035(2)&   1.014(74)&    2.67\\
  0.10&   5.034(3)&   1.024(90)&    2.98\\
  0.15&   5.038(1)&   0.954(30)&    0.49\\
  0.175&   5.036(3)&   0.984(77)&    3.49\\
  0.20&   5.036(3)&   0.977(77)&    3.33\\
  0.22&   5.037(3)&   0.926(117)&    3.09\\
\end{tabular}
\end{table}

\begin{table}
\setlength{\tabcolsep}{1.1pc}
\caption{Results of the least squares quadratic fits to the critical line}
\vspace{0.3cm}
\label{tablefitcrit}
\begin{tabular}{r r r r r r}
 & $\mu$ range & a & b & SSR & $\sum \Delta \beta ^2$ \\
\hline
\\
p1 & [0:0.17] & 5.039(2) & 0.79(12) &1.4e-05 & 8 e-06  \\
p2 & [0:0.20] &5.038(2) &0.90(10) &2.8e-05 &1.0 e-05  \\
p3 & [0:0.22] &5.038(2) & 0.94(7) &3.1e-05 &1.2 e-05 \\
\end{tabular}

\end{table}

\begin{table}
\caption{Fourier coefficients for the chiral condensate. 
The constant and the first coefficeint are satisfactory
determined for $T < T_E$}
\begin{tabular}{l l l l l}
$\beta$ & $a_F^{0}$ & $a_F^{1}$ & $a_F^{2}$ & $\chi^2 /d.o.f.$ \\
\hline
5.03 &  0.931(1)  &  -0.0251(22) & 0(0) &  1.31 \\
5.03 &  0.930(2) &  -0.0230(26) &  -0.0034(26) &   1.15 \\
\hline
5.01  &     0.974(1) &  -0.0106(9) & 0(0) & 0.87 \\
5.01 &  0.974 (1) &  -0.0107 (10) &-0.0003  (12)&  0.98 \\
\hline
5.10 & 0.396 (1) &  -0.024 (1) & 0(0)    & 1.22294 \\
5.10 & 0.404(4) & -0.38(6) & 0.007(3) & 0.96 \\
\end{tabular}
\label{cosfitpbp}
\end{table}

\begin{table}
\caption{Coefficients of a polynomial fit 
for the chiral condensate f(x) = $a^P_0 - 72a^P_2x^2 + 864a^P_4x^4$.}
\begin{tabular}{l l l l l}
$\beta$ & $a^P_{0}$ & $a^P_{2}$ & $a^P_{4}$ & $\chi^2 /d.o.f.$ \\
\hline
5.03 &  0.905(2)  &  -0.0272(40) & -0.0261(70) &  1.17 \\
5.01 &  0.963(1) &  -0.011(3) & -0.009(4) & 1.03 \\
5.10 & 0.372 (1) &  -0.019 (1) & 0(0)    & 1.03 \\
\end{tabular}
\label{polfitpbp}
\end{table}

\newpage

\newpage

\newpage

\newpage

\newpage
\begin{figure}
\vspace{2cm}
\include{reale_uno}
\label{fig:reale_uno}
\end{figure}

\newpage
\begin{figure}
\vspace{2cm}
\include{reale_due}
\label{fig:reale_due}
\end{figure}

\newpage
\begin{figure}
\vspace{2cm}
\include{reale_tre}
\label{fig:reale_tre}
\end{figure}

\newpage
\begin{figure}
\vspace{2cm}
\include{reale_quattro}
\label{fig:reale_quattro}
\end{figure}

\newpage
\begin{figure}
\vspace{2cm}
\include{reale_cinque}
\label{fig:reale_cinque}
\end{figure}

\end{document}